\documentclass{article}

\usepackage{arxiv}

\usepackage[utf8]{inputenc} 
\usepackage[T1]{fontenc}    
\usepackage{hyperref}       
\usepackage{url}            
\usepackage{booktabs}       
\usepackage{amsmath,amssymb}       
\usepackage{nicefrac}       
\usepackage{microtype}      
\usepackage{lipsum}
\usepackage{multirow,bm}
\usepackage{graphicx}
\usepackage{geometry}
\usepackage{array}
\usepackage{algorithm,algorithmic}
\usepackage{xcolor}
\graphicspath{ {./images/} }

\usepackage{titlesec}
\titleformat{\paragraph}
{\normalfont\normalsize\bfseries}{\theparagraph}{1em}{}
\titlespacing*{\paragraph}
{0pt}{3.25ex plus 1ex minus .2ex}{1.5ex plus .2ex}

\title{Reduced Order Probabilistic Emulation for Physics-Based Thermosphere Models}

\author{
 Richard J. Licata \\
  Dept. of Mechanical and Aerospace Engineering \\
  West Virginia University\\
  Morgantown, WV 26505 \\
  \texttt{rjlicata@mix.wvu.edu} \\
   \And
 Piyush M. Mehta \\
  Dept. of Mechanical and Aerospace Engineering \\
  West Virginia University \\
  Morgantown, WV \\
}

\begin{document}
\maketitle
\begin{abstract}

The geospace environment is volatile and highly driven. Space weather has effects on Earth's magnetosphere that cause a dynamic and enigmatic response in the thermosphere, particularly on the evolution of neutral mass density. Many models exist that use space weather drivers to produce a density response, but these models are typically computationally expensive or inaccurate for certain space weather conditions. In response, this work aims to employ a probabilistic machine learning (ML) method to create an efficient surrogate for the Thermosphere Ionosphere Electrodynamics General Circulation Model (TIE-GCM), a physics-based thermosphere model. Our method leverages principal component analysis to reduce the dimensionality of TIE-GCM and recurrent neural networks to model the dynamic behavior of the thermosphere much quicker than the numerical model. The newly developed reduced order probabilistic emulator (ROPE) uses Long-Short Term Memory neural networks to perform time-series forecasting in the reduced state and provide distributions for future density. We show that across the available data, TIE-GCM ROPE has similar error to previous linear approaches while improving storm-time modeling. We also conduct a satellite propagation study for the significant November 2003 storm which shows that TIE-GCM ROPE can capture the position resulting from TIE-GCM density with < 5 km bias. Simultaneously, linear approaches provide point estimates that can result in biases of 7 -- 18 km.

\end{abstract}


\section{Introduction}\label{sec:intro}

Over the past decade, private companies have been assembling megaconstellations in low Eath orbit (LEO) increasing traffic in the crowded orbital regime. Meanwhile, accurate specification of the thermosphere continues to be a major challenge for the modeling community. This combination creates an environment where the future position of thousands of valuable objects are considerably uncertain. Atmospheric drag is the major contributor to this uncertainty, and thermospheric density remains one of the prominent drivers behind this.

Thermospheric density varies with solar irradiance which changes over the Sun's 11-year solar cycle in addition to cycling with the Sun's 27-day rotational period \cite{solar_cycle_rotation}. However, this is a general trend since density can change from other factors such as composition changes and geomagnetic storms \cite{Comp,current_state}. One origin for geomagnetic storms is the presence of an Earth-facing coronal hole. The perturbations from the resulting fast solar winds tend to occur with a regular frequency of 27 days during low solar activity, because the low density regions on the Sun evolve slowly \cite{coronal_holes}. Conversely, coronal mass ejections (CMEs) are less regular and cause much larger and more sudden increases in thermospheric density. CMEs can cause the thermosphere to heat and cool rapidly and are a source of higher errors with current thermosphere models \cite{heating_cooling,therm_storm}. This is largely due to  a severe underrepresentation of strong geomagnetic storms in current datasets due to the rarity of these events \cite{storm_drag}. 

High-latitude forcing from geomagnetic storms is improved in physics-based models, and they are often used to study the thermosphere during storms \cite{storm1,storm2,storm3}. Physics-based models numerically solve the equations that govern the thermosphere, typically using finite-differencing or spectral methods \cite{Emmert07}. Predictions from these models are invaluable, because they represent a physical evolution of the system. However, their computational expense creates limitations for their use in an operational setting. More efficient thermosphere modeling has been gaining traction over recent years with the use of reduced order models (ROMs). Linear ROMs have been developed using Dynamic Mode Decomposition (DMD) based on popular empirical and physics-based models \cite{MehtaROM,TLE_ROM,TLE_ROM2}. Operating in a reduced space also makes data assimilation more feasible for high-dimensional systems \cite{MehtaDACal}. Machine learning (ML) provides an avenue for nonlinear ROM development and uncertainty quantification (UQ) and has been explored for thermosphere modeling \cite{HASDM_ML,UQtech}.

In this work, we build on previous nonlinear thermosphere ROMs by developing a reduced order probabilistic emulator for the Thermosphere Ionosphere Electrodynamics General Circulation Model (TIE-GCM) using ensemble recurrent neural networks (RNNs) \cite{TIEGCM}. This is an improvement over previous work through the incorporation of nonlinear modeling and the capability to provide uncertainty distributions for TIE-GCM density predictions. We start by describing the TIE-GCM dataset, ML modeling approach, and DMD (for comparison). We then show results for both five-day and year-long dynamic predictions using DMD and the ML ensemble model. Finally, we show the operational implications of having a probabilistic dynamic ROM by considering a satellite orbit using different modeling approaches.

\section{Data and Methods}\label{sec:method}

Developing effective machine learning models is no simple task. In order to train a model to high accuracy, the dataset has to be properly chosen and refined. Users must be aware of the input/output (I/O) relationship of the system in order to provide the model with sufficient information to properly understand the system. Furthermore, proper metrics need to be established to determine accuracy of the model and identify any possible areas for improvement.

\subsection{TIE-GCM Data}

The Thermoshere-Ionosphere- Electrodynamics General Circulation Model is part of the Thermosphere General Circulation Model (TGCM) series dating back to 1981 \cite{TGCM}. TIE-GCM was developed by Richmond et al. \cite{TIEGCM_orig} and managed to incorporate electrodynamic interactions between the thermosphere and ionosphere systems.  TIE-GCM is a predictive space weather model that starts with an initialized state of the thermosphere and dynamically evolves that system, through finite differencing methods, based on a set of inputs \cite{TIEGCM}. TIE-GCM is important for studying different phenomena in the thermosphere due to its ability to provide a physically-constrained representation of the system. However, the high computational cost and need for significant parallelization limits its application to operations or collision assessment. 

For this work, the data comes from TIE-GCM for solar cycle 23 (1996--2008) along with an additional year of simulated input drivers \cite{ROM}. The year of simulated inputs, called “Sim1”, allows for a much smaller dataset to contain many geomagnetic storms with varying levels of solar ($F_{10}$) and geomagnetic activity ($Kp$). Geomagnetic storms are infrequent, especially strong storms, but they are a vital aspect of modeling density for both scientific research and operational use. The high frequency of storms and rapidly changing solar activity in Sim1 is useful for analysis to determine overall performance in a manageable number of samples.

This dataset has the following spatial resolution: 24 local solar time values, 20 latitude values, and 16 altitudes evenly spaced between 100 and 450 km. The spatial dimensionality of this dataset is too high for uncertainty quantification methods and observability for data assimilation \cite{MehtaROMCal}. Therefore, principal component analysis (PCA) is applied to reduce the dataset. The PCA coefficients ($\alpha_i$) are derived through, 
\begin{equation} \label{eq1}
\begin{split}
\mathbf{x}\left(\mathbf{s},t\right)=\mathbf{\bar{x}}\left(\mathbf{s}\right)+\mathbf{\widetilde{x}}\left(\mathbf{s},t\right)
\;\;\;\textrm{and}\;\;\;
\mathbf{\widetilde{x}}\left(\mathbf{s},t\right)=\sum^r_{i=1}\alpha_i\left(t\right)U_i\left(s\right)
\end{split}
\end{equation}
where $\mathbf{x}\in\mathbb{R}^m$ is the model output state (TIE-GCM mass density at all $m$ = 7,680 grid locations) after a \textit{log\textsubscript{10}} transformation, $\mathbf{\bar{x}}$ is its mean, $\mathbf{\widetilde{x}}$ is its deviation from the mean, $\mathbf{s}$ represents the spatial dimension, $r$ is the user-selected truncation order (\textit{r}=10), and $U_i$ are orthogonal modes, or basis functions. This can be achieved using the \textit{svds} function in \textit{MATLAB} or the \textit{PCA} function in the \textit{sci-kit learn decomposition} module for \textit{Python} \cite{MehtaROM,MehtaROMCal,scikit-learn}.

\begin{figure}[htb!]
	\centering
	\small
	\includegraphics[width=0.98\textwidth]{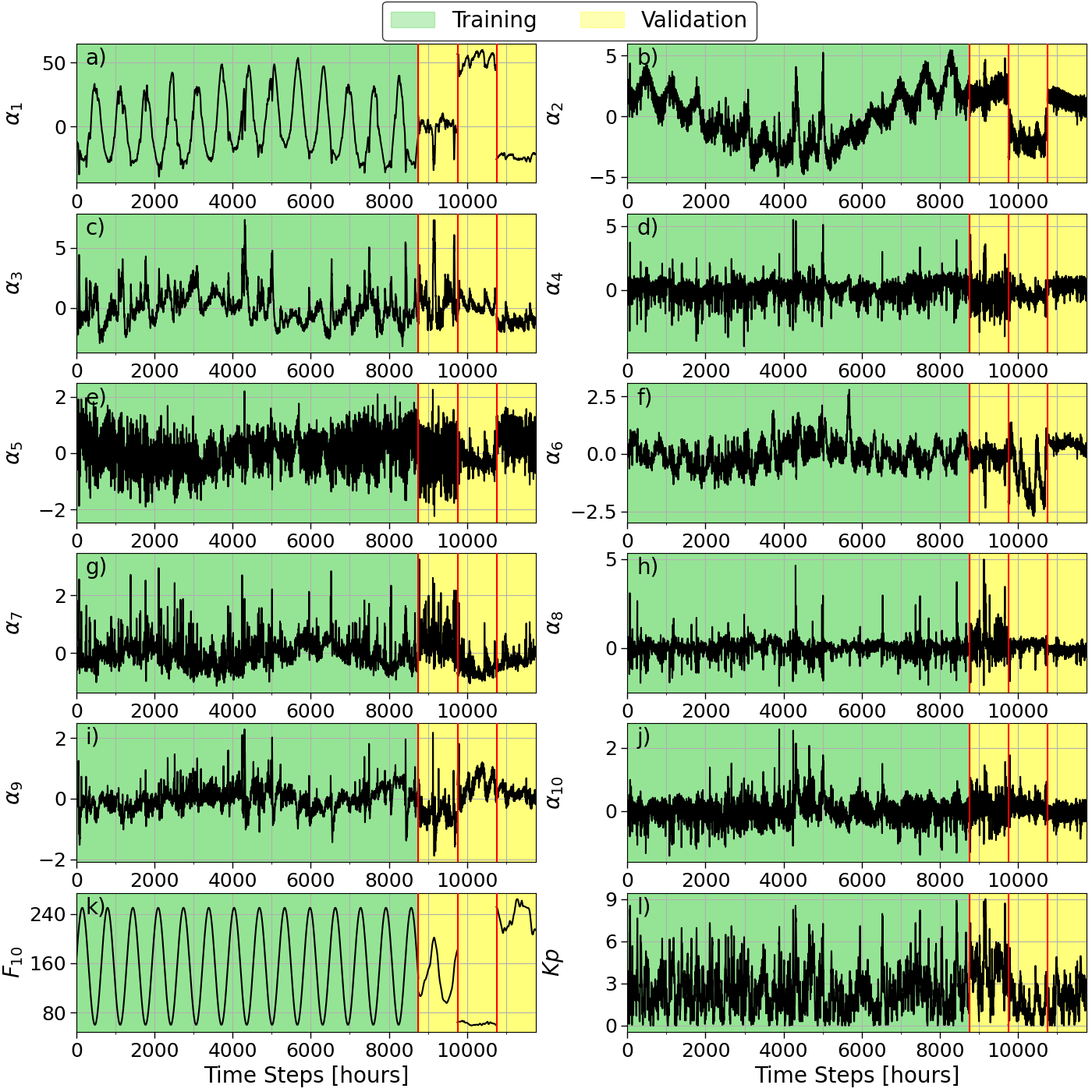}
	\caption{TIE-GCM PCA coefficients for Sim1 dataset and selected validation segments (a -- j) with corresponding \textit{F\textsubscript{10}} (k) and \textit{Kp} (l). The validation segments are shown as presented in the text (late 2003, mid-2008, early 2002).}
	\label{f:LSTM_Sim1_Tuner}
\end{figure}

For the tuner (described in Section \ref{sec:UQ}), the Sim1 dataset was chosen for manageable run times, and three segments from the solar cycle were chosen for validation. The three 1,000 sample segments represent high geomagnetic activity (late 2003), solar minimum (mid-2008), and solar maximum (early 2002). The PCA coefficients and space weather drivers for this tuning and testing dataset are shown in Figure \ref{f:LSTM_Sim1_Tuner}.

The models developed based on this dataset use six inputs: \textit{F\textsubscript{10}}, \textit{Kp}, and sinusoidal transformations for universal time (\textit{UT}) and day of year (\textit{doy}). These inputs are referred to as $t_1$--$t_4$. 
\begin{equation} \label{eqT}
\begin{split}
t_1=sin\left(\frac{2\pi doy}{365.25}\right),\;\;\;\;
t_2=cos\left(\frac{2\pi doy}{365.25}\right),\;\;\;\;
t_3=sin\left(\frac{2\pi UT}{24}\right),\;\;\;\;
t_4=cos\left(\frac{2\pi UT}{24}\right)
\end{split}
\end{equation}
These transformations account for the seasonal and diurnal variations in the state of the thermosphere while remaining continuous about the boundaries. The inputs and outputs are also normalized using standard normalization, given as
\begin{equation} \label{eqstdnorm}
\begin{split}
\widetilde{\theta} = \frac{\theta-\textrm{mean}(\theta)}{\textrm{standard deviation}(\theta)}
\end{split}
\end{equation}
where $\theta$ is an input/output variable, and $\widetilde{\theta}$ is the normalized $\theta$. This ensures that all all input and output variables to the model will have the same first two moments as a standard normal distribution. This operation can be easily reversed to obtain un-normalized model predictions.

\subsubsection{Model Development}\label{sec:LSTM_model development}

TIE-GCM is a dynamic model, meaning it models the temporal evolution of the system. This aspect of TIE-GCM makes it valuable in scientific studies and therefore a surrogate model should work the same way. To develop a dynamic ML model, we leverage Long-Short Term Memory neural networks (LSTMs). Like vanilla recurrent neural networks, a number of "lag steps" must be defined. For TIE-GCM ROPE, three lag steps are used. Since the TIE-GCM dataset has a cadence of one-hour, this corresponds to a three-hour window which is generally enough time for perturbations to the input to be seen throughout the thermosphere. The internal cell memory allows for the longer-term effects to be accounted for. Licata et al. \cite{scienceML} found that the data-driven CHAMP-ML model had the strongest relationship between density and either current geomagnetic indices or indices from the last three hours. 

\subsection{Long Short-Term Memory Neural Networks}\label{sec:RNN}

Long-Short Term Memory neural networks are a type of recurrent neural network that can reliably keep track of long-term historical information, making it suitable for SW applications where it is important to capture the hysteric property of the system. During training, the LSTM learns not only the short term trends and how to predict data, but also the relevance of previous information even with large time lags \cite{Hochreiter}. This makes them versatile and capable of picking up dynamic trends. In this application, the LSTM can “subconsciously” acquire knowledge on the underlying physics allowing it to replicate the performance of the high-fidelity physics model.

\begin{figure}[htb!]
	\centering
	\small
	\includegraphics[width=0.90\textwidth]{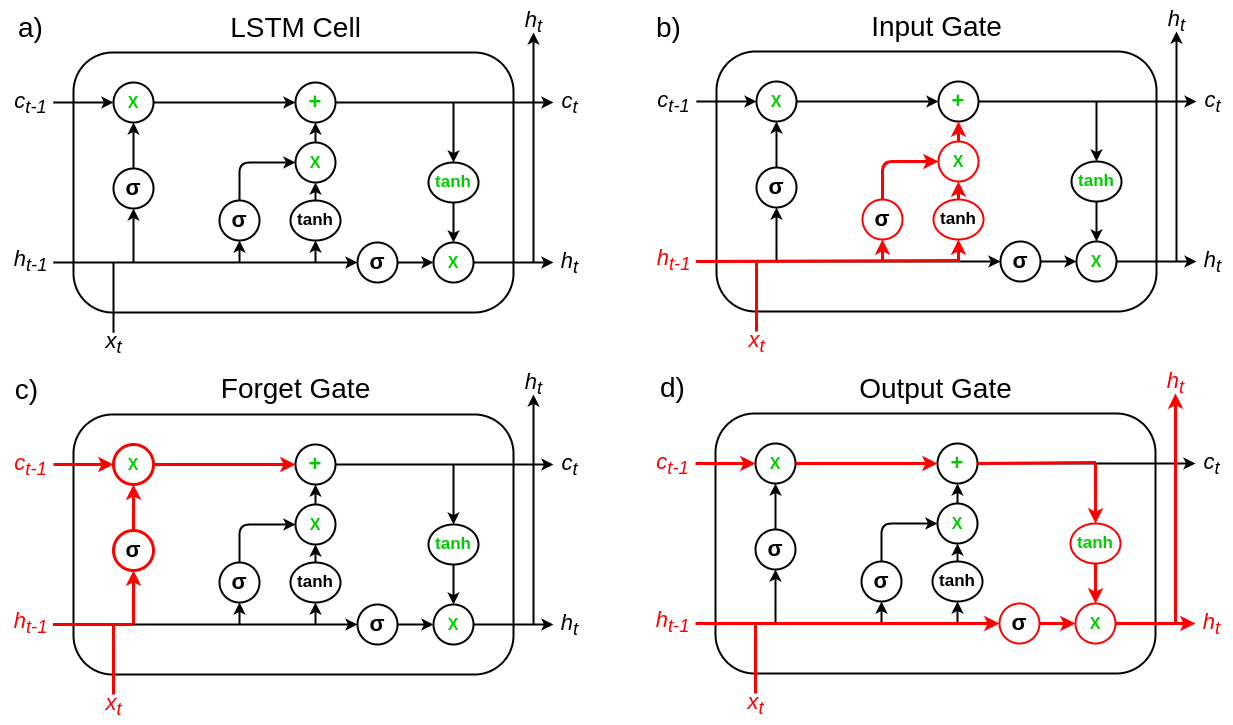}
	\caption{Overall construction of the LSTM cell (a) with the input gate (b), forget gate (c), and output gate (d) highlighted in red. Green is used to denote point-wise operations.}
	\label{f:lstmcell}
\end{figure}

For recurrent neural networks like an LSTM, the corresponding inputs and outputs are concatenated, and the inputs for a given time-step are these stacked inputs and output combination. The number of previous time-steps used is a new hyperparameter when dealing with RNNs. LSTMs are dynamic neural networks that deviate from traditional RNNs through their use of an internal cell state, specifically its input, forget, and output gates \cite{Hochreiter} (see Figure \ref{f:lstmcell}).

In Figure \ref{f:lstmcell}, \textit{x} refers to the input to the cell, \textit{c} refers to the cell's internal state/memory, and \textit{h} is the output. \textit{t} and \textit{t-1} denote the current and previous step, respectively. The three $\sigma$ nodes refer to internal layers with the sigmoid activation function, and tanh refers to either a layer with the tanh activation function or a point-wise operation -- dependant on the color. The LSTM cell (a) is shown with each aforementioned gate highlighted: input gate (b), forget gate (c), and output gate (d). 

The internal sigmoid layers function similar to typical binary gates. If a gate is open (1), information passes through. Conversely if a gate is closed (0), no information gets through. As sigmoid has a continuous range between zero and one, it is ideal to function as a gate while keeping the LSTM cell differentiable. For the input gate, the input and previous output information get passed both the the sigmoid and tanh layers. The tanh layer acts as a normal neural network layer, while the sigmoid layer determines how much out the tanh output passes through. The forget gate passes the input and previous output information through another sigmoid layer that will interact with the previous cell state. The output of the sigmoid layer will determine how much of the previous cell state will pass through. 

The last gate is the output gate. This uses the input and previous output information to determine how much of the output will pass through to the next layer. While only certain parts are highlighted in (d), the output gate affects information from the other two gates. While these three gates are described independently, the information passes through the cell concurrently. The internal cell state is updated, and the information passes onto the downstream layer. The internal cell state is how LSTMs can keep track of long-term information.

\subsubsection{Data Preparation} \label{sec:lstm_data_prep}

In standard feed-forward neural networks, inputs and outputs simply need to have the same length about the first axis to achieve supervised training. However, recurrent neural networks require further processing. Consider the number of inputs ($n_{inp}$) and number of outputs ($n_{out}$) for a given dataset with $n$ samples. The concatenation will result in an array of shape $n \times (n_{out}+n_{inp})$. A new hyperparameter for RNNs is the number of lag-steps ($n_{LS}$). This is the number of previous time steps the model will use to make a single prediction (think short-term memory for an LSTM). 

The data must be stacked, so each row contains outputs and inputs for each lag-step and the current step. This orders in least-to-most recent from left-to-right. The data will now be of the shape $n \times (n_{LS}+1)(n_{out}+n_{inp})$. The last $n_{inp}$ columns are then dropped as they are not needed. The data can be split into training inputs and outputs where the first $n \times n_{LS}(n_{out}+n_{inp})$ columns are inputs and the last $n_{out}$ columns are the associated output. The final step is to reshape the input data to the shape $n \times n_{LS} \times (n_{out}+n_{inp})$.

For our model development, we chose 2002--2008 for training (61,368 samples) due to its containment of the high and low extreme \textit{F\textsubscript{10}} values for the solar cycle. This leaves mid-1996 through 2001 for testing (50,688 samples). For the validation set, we chose a subset of the Sim1 dataset with 1,250 samples (approximately 52 days). This period contains multiple storms and a wide range of \textit{F\textsubscript{10}}.

\subsubsection{Training and Evaluation}\label{sec:LSTM_train_eval}

With the LSTM training data prepared for the supervised learning task, a model can be trained. Typical LSTM training involves one-step prediction. This means that for each time step, the model uses the last $n_{LS}$ sets of true inputs and outputs to make the next prediction. As it goes through an epoch sequentially, the model will not only be updating weights in the way described for feedforward neural networks, it also updates the weights associated with the three sigmoid layers and one tanh layer within each LSTM cell. Between epochs, the LSTM internal state $c$ will be changing as information passes through each cell.

\begin{figure}[htb!]
	\centering
	\small
	\includegraphics[width=\textwidth]{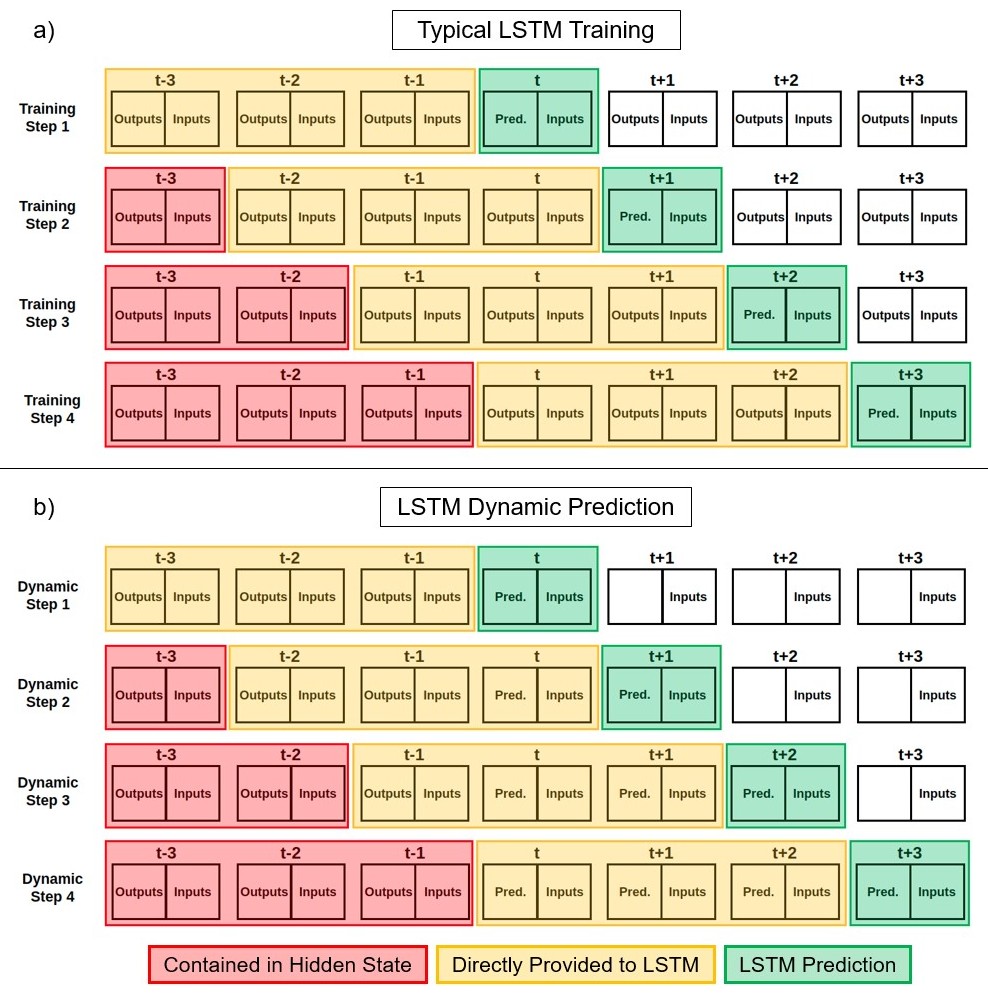}
	\caption{Steps for typical LSTM training (a) and steps for an iterative dynamic prediction (b) with $n_{LS}=3$. Although the predictions and inputs for the evaluation step are highlighted in green, the model only predicts the output.}
	\label{f:lstm_train_pred}
\end{figure}

LSTMs also deal with a concept known as resetting the internal cell state. When reaching a temporal discontinuity, it is important to wipe the internal memory of every LSTM cell. Otherwise, it will be using irrelevant information to make predictions. This is typically done at the end of an epoch when it reaches the end of the training time period. Resetting the state is also critical when evaluating the model on different time periods. The approach for one-step LSTM training is shown in Figure \ref{f:lstm_train_pred}, panel (a).

It is important to note that in the training phase, the predictions are ignored -- for lack of a better term -- when moving to the next time-step. The prediction is replaced with the true output for that time period. However, this is not realistic in a forecasting environment. If a forecast is required for the next 25 steps (consider $n_{LS}=3$ from Figure \ref{f:lstm_train_pred}), you would only have data up to the previous step \textit{t-1}. The LSTM will make a prediction for time \textit{t} and has to use that to make its prediction for time \textit{t+1}. At this point, the model is taking two true values and one predicted value to make the next prediction. At \textit{t+2}, there is only one true value and two predicted values. From \textit{t+3} to the end of the prediction, the model will use solely the inputs and previous predictions. This is referred to as a dynamic prediction and is shown in Figure \ref{f:lstm_train_pred} panel (b).

\subsection{Uncertainty Quantification}\label{sec:UQ}

For other models developed by the authors, direct probability distribution prediction resulted in both low error and robust uncertainty estimates \cite{UQtech,msis_uq}. However, they were static, not recurrent, neural networks. Early efforts showed that the direct probability method was strongly underestimating uncertainty due to the one-step training and dynamic prediction process. To attempt to combat this, we overhauled the default LSTM training process and force it to use a dynamic training process, mimicking its operational usage to get better predictions of $\sigma$. This required batch averaging and resulted in poor mean prediction capabilities. The final approach was to develop $\mu$ and $\sigma$ models separately to try and leverage the benefits of both previous tests. The $\mu$ models could perform dynamic prediction with low error, but the $\sigma$ models still struggled to provide meaningful uncertainty estimates. Another approach to uncertainty quantification, popular in terrestrial and space weather applications, is ensemble modeling \cite{Dst_ensemble,storm_ensemble}. We therefore develop many individual LSTMs and combine the results in such a way to obtain accurate predictions and reliable uncertainty estimates.

Hyperparameter tuners were run on the Sim1 dataset to determine architectures for the aforementioned direct probability tests. While those did not yield an adequate final model, the last attempt (separate $\mu$ and $\sigma$ modeling approach) provided architectures for potential ensemble models. The tuner options and search space used for the MSE models are provided in Table \ref{t:LSTM_KT}. These models were evaluated on the validation set shown earlier in Figure \ref{f:LSTM_Sim1_Tuner}, and the top two architectures were used for future training.

\begin{table}[htb!]
	\fontsize{10}{10}\selectfont
    \caption{Hyperparameter tuner parameters (left) and search space (right) for the mean square error LSTM.}
   \label{t:LSTM_KT}
        \centering 
   \begin{tabular}{| c | c | c | c|} 
      \hline 
            \textbf{Tuner Option} & \textbf{Choice} & \textbf{Parameter} & \textbf{Values/Range} \\ \hline 
            \multirow{2}{*}{\textit{Scheme}} & Bayesian & \textit{Number of} & \multirow{2}{*}{1--3}\\ 
            & Optimization & \textit{LSTM Layers} & \\ \hline 
            \multirow{2}{*}{\textit{Total Trials}} & \multirow{2}{*}{50} & \multirow{2}{*}{\textit{LSTM Neurons}} & min = 32, max = 512,\\
            & & & step = 4 \\\hline 
            \multirow{2}{*}{\textit{Initial Points}} & \multirow{2}{*}{25} & \multirow{2}{*}{\textit{LSTM Activations}} & \multirow{2}{*}{tanh, sigmoid, softsign} \\ 
            & & & \\ \hline
            \multirow{2}{*}{\textit{Repeats per Trial}} & \multirow{2}{*}{3} & \textit{Number of} & \multirow{2}{*}{1--3} \\ 
            & & \textit{Dense Layers} & \\ \hline
            \textit{Minimization} & \multirow{2}{*}{val\_loss} & \multirow{2}{*}{\textit{Dense Neurons}} & min = 64, max = 1024,\\
            \textit{Parameter} & & & step = 4 \\\hline 
            \multirow{2}{*}{\textit{Epochs}} & \multirow{2}{*}{2,500} & \multirow{2}{*}{\textit{Dense Activations}} & tanh, sigmoid, softsign, \\ 
            & & & relu, elu, softplus\\ \hline
            \textit{Early Stopping} & \multirow{2}{*}{val\_loss} & \multirow{2}{*}{\textit{Dense Dropout}} & min = 0.01, max = 0.50,\\ 
            \textit{Criteria} & & & step = 0.01 \\ \hline
            \textit{Early Stopping} & \multirow{2}{*}{75 epochs} & \multirow{2}{*}{\textit{Optimizer}} & RMSprop, Adam,\\
            \textit{Patience} & & & Adadelta, Adagrad \\
      \hline
   \end{tabular}
\end{table}

Early modeling efforts showed that the Sim1 dataset, while valuable for testing and tuning, did not provide adequate model performance when used for training. Therefore, the training data for the final model comes from the TIE-GCM outputs spanning 2002--2008. This provides the model with the highest and lowest extremes of solar activity seen in the solar cycle. PCA is performed again on this time period, and all density data (1996--2008 and Sim1) is transformed using the basis functions from this training set. For validation, a block of 1,250 samples (approximately 52 days) from Sim1 is used as it provides a wide range of conditions for evaluation in a short period.

One potential problem with this data is that there must be continuity for the LSTM internal state, but the model should not see the data in the same order every epoch -- starting with solar maximum and ending with solar minimum. To avoid potential issues, the seven years are split into 490 segments with 125 time steps (approximately five days) of continuous data within them. The training process can be modified such that the model can be trained on each 125 sample continuous segment, and the internal state can be reset after each one. These 490 segments can be shuffled such that the LSTM sees different ordering of the data while being able to fine-tune its internal cell parameters without the threat of discontinuous data. 

Another stark difference between Sim1 and 2002--2008 is the low relative frequency of geomagnetic storms in the historical period. Early testing also showed that the LSTMs trained on the historical data were more accurate overall compared to Sim1 models, but they had higher storm-time errors. A straightforward solution is to use sample weighting: applying an importance to each individual sample. We use a simple algorithm based on the frequency of samples at different \textit{Kp} levels such that the number of samples within a \textit{Kp} bin multiplied by the number of samples in that bin is equal across all bins. This can help enforce importance to storm-time samples based on the relative frequency of these events.

At this point, the architectures for both models are finalized. To obtain the final $\mu$ model, it is trained on this new dataset with the described sample weighting scheme. At first, the model is trained using a typical one-step training method, but the loss is averaged over the 125 sample segment. This is performed for up to 2,500 epochs with early stopping based on mean absolute error in the density space for the validation set. After each validation segment, the true and predicted PCA coefficients are converted back to density through the inverse PCA transformation. Since the importance of the coefficients are not uniform, the MSE in the PCA space is not directly correlated to the best model. After the model is finished with batch training, it continues training without batch averaging using a smaller learning rate until the early stopping criteria is met once more. We obtain five models using this approach for the best two architectures resulting in ten total models to make up the ensemble.

\subsubsection{Weighted Averaging and Uncertainty Scaling}\label{sec:ensemble_scaling}

To derive the weighting arrays and uncertainty scaling factors, we do all computations separately within each architecture ($i=1,2$). The combination of the two architectures completes the ensemble, predicting different possibilities for a period of interest. Each individual model uses its own outputs as inputs (dynamic prediction), so the combination is done post-prediction. 

The predictions of the PCA coefficients from each model will differ with varying levels of accuracy. Instead of simply averaging the predictions across the five models (for a given architecture), we opt to determine weighting factors based on the relative error of each model. To achieve this, each model is evaluated across the training set in five-day dynamic segments. The predictions of the PCA coefficients for each model and period are saved for later evaluation. The mean absolute error is computed for each model and each coefficient over the entire training set resulting in a 5$\times$10 array. Using this, the weights ($w$) are computed as,
\begin{equation} \label{eqweights}
\begin{split}
w_{i,j,k} = \frac{\widetilde{w}_{i,j,k}}{\sum^{10}_{j=1} \widetilde{w}_{i,j,k}} \;\;\;\;\;\;\;\;\; where \;\;\;\;\;\;\;\;\; \widetilde{w}_{i,j,k} = \frac{1}{\textrm{MAE}_{i,j,k}} 
\end{split}
\end{equation}
where $i$, $j$, and $k$ refer to each architecture, model, and PCA coefficient, respectively.  $\widetilde{w}_{i,j,k}$ denotes the weights at an intermediate step before normalization. Once computed, the weighted mean and variance for each PCA coefficient from each architecture can be obtained,
\begin{equation} \label{eqweights2}
\begin{split}
\hat{\alpha}_{i,k,t} = \sum^{5}_{j=1} \;\;\; w_{i,j}\hat{\alpha}_{i,j,k,t} \;\;\;\;\;\;\;\;\; and \;\;\;\;\;\;\;\;\; \hat{\sigma}_{i,k,t}^2 = \sum^{5}_{j=1}w_{i,j}\left(\hat{\alpha}_{i,k,t}-\hat{\alpha}_{i,j,k,t}\right)^2
\end{split}
\end{equation}
where $\hat{\alpha}_{i,k,t}$ and $\hat{\sigma}_{i,k,t}^2$ are the ensemble mean and variance for the \textit{i\textsuperscript{th}} architecture and \textit{k\textsuperscript{th}} PCA coefficient at time $t$, respectively. $\hat{\alpha}_{i,j,k,t}$ refers to the corresponding prediction from each of the five models. While this will provide a distribution under a Gaussian assumption, it does not guarantee robustness and reliability of the resulting uncertainty estimates. This can be improved with so-called $\sigma$ scaling. Laves et al. \cite{sigmascaling} came up with a scaling factor (\textit{s}) to scale model $\sigma$ to better represent uncertainty. The scaling factor can be computed using the following equation, based on Equation 9 in \cite{sigmascaling}.
\begin{equation} \label{eqsigmascale}
\begin{split}
S_{i,k} = \sqrt{\frac{1}{n_{tr}}\sum^{n_{tr}}_{t=1}\frac{\left(\alpha_{k,t}-\hat{\alpha}_{i,k,t}\right)^2}{\hat{\sigma}_{i,k,t}^2}} 
\end{split}
\end{equation}
In Equation \ref{eqsigmascale}, $S_{i,k}$ is the scaling factor for the \textit{i\textsuperscript{th}} architecture and \textit{k\textsuperscript{th}} PCA coefficient, \textit{n\textsubscript{tr}} is the number of time-steps in the training set, and $\alpha_{k,t}$ is the true/reference value for the \textit{k\textsuperscript{th}} PCA coefficient at time $t$. The ensemble weights and scaling factors are saved for all later model use. The overall ensemble mean and variance can be computed as,
\begin{equation} \label{eqcombination}
\begin{split}
\hat{\alpha}_{k,t} = \frac{1}{2} \sum^2_{i=1}\hat{\alpha}_{i,k,t} \;\;\;\;\;\;\;\;\; and \;\;\;\;\;\;\;\;\; \hat{\sigma}^2_{k,t} = \frac{(n_1-1)\hat{\sigma}_{1,k,t}^2 + (n_2-1)\hat{\sigma}_{2,k,t}^2}{n_1+n_2-2} = \frac{\hat{\sigma}_{1,k,t}^2+\hat{\sigma}_{2,k,t}^2}{2}
\end{split}
\end{equation}
where $n_1$ and $n_2$ are the number of models within each architecture. This is the pooled variance formula which reduces to a simple average due to an equal number of models in each architecture. The final ensemble will be referred to as TIE-GCM reduced order probabilistic emulator (ROPE) due to its ability to provide Gaussian uncertainties and to function as a reduced order emulator for TIE-GCM (see Section \ref{sec:LSTMemulation}).

\subsection{Metrics}\label{sec:metrics}

To assess the general predictive performance of TIE-GCM ROPE, we use the mean absolute error in terms of mass density. This is shown in a relative form as,
\begin{equation} \label{eqmae}
\begin{split}
\textrm{MAE} = \frac{100\%}{n \cdot m}\sum^n_{t=1} \sum^m_{s=1} \frac{\left|\rho_{s,t}-\hat{\rho}_{s,t}\right|}{\rho_{s,t}}
\end{split}
\end{equation}
where \textit{n} and \textit{m} are the number of time-steps and spatial locations, respectively. $\rho$ and $\hat{\rho}$ refer to the TIE-GCM and predicted density, respectively. This metric is used for consistency with previous density modeling efforts by the authors \cite{HASDM_ML,UQtech,msis_uq}. Previous work also used the calibration error score (CES) metric to determine how robust and reliable a model's uncertainty estimates are,
\begin{equation} \label{eqces}
\begin{split}
\textrm{CES} = \frac{100\%}{r \cdot n_{PI}}\sum^r_{k=1} \sum^{n_{PI}}_{u=1} \Big|p(\alpha_{k,u})-p(\hat{\alpha}_{k,u})\Big|
\end{split}
\end{equation}
In Equation \ref{eqces}, $n_{PI}$ is the number of prediction intervals of interest, $p(\alpha_{k,u})$ is the expected cumulative probability (e.g. 0.95), and $p(\hat{\alpha}_{k,u})$ is the observed cumulative probability or the percentage of samples that are within the bounds. The prediction intervals used to compute CES in this work range from 5\% to 99\% in 5\% intervals. More details on the calculation of this metric can be found in Section 2.5.1 from \cite{HASDM_ML}.

\subsection{DMDc Approaches}\label{sec:DMDtest}

Over the last several years, some researchers have developed dynamic reduced order models (ROMs) for empirical and physics-based thermosphere models alike. Mehta et al. \cite{MehtaROM} used PCA on TIE-GCM data and developed a dynamic ROM using dynamic mode decomposition (DMD) with control (or DMDc). This approach has been applied by Gondelach and Linares \cite{TLE_ROM,TLE_ROM2} on the NRLMSISE-00, JB2008, and TIE-GCM models with the goal of data assimilation. DMDc is based on the assumption of the linear relationship between successive time steps and the processes that drive the system,
\begin{equation} \label{eqDMD1}
\begin{split}
\mathbf{x}_{k+1} = \mathbf{Ax}_k + \mathbf{Bu}_k
\end{split}
\end{equation}
where $\mathbf{x}$ denotes the state, $\mathbf{A}$ is the dynamic matrix, and $\mathbf{B}$ is the input matrix relating the system inputs/drivers and successive state of the system. Here, $k$ refers to the time step. In a matrix format, this is achieved by $\mathbf{x}_{k+1}$ being the PCA coefficients from $2:n$, $\mathbf{x}_k$ being the coefficients from $1:n-1$, and $\mathbf{u}_k$ being the chosen drivers from $1:n-1$. As this is used as a benchmark for ML dynamic model development, the reader is referred to Proctor et al. \cite{DMDc} for the theory behind DMDc and to Mehta et al. \cite{MehtaROM} for details on its application to this dataset.

Given that considerable work has been done by other researchers for dynamic thermosphere modeling with DMDc, we use it as a baseline to compare with the LSTM. Gondelach and Linares \cite{TLE_ROM} compared DMDc for NRLMSISE-00, JB2008, and TIE-GCM using linear and nonlinear inputs. For TIE-GCM, they used \textit{Kp\textsuperscript{2}} and \textit{Kp}$\cdot$\textit{F\textsubscript{10}} to try and overcome DMDc's limitation of linearity. To conduct a thorough test, we consider the LSTM test set of 1996--2001, and split the data into three segments using five-day dynamic prediction windows. (1) \textit{Kp\textsubscript{max}} $<5$, (2) $5\leq$ \textit{Kp\textsubscript{max}} $<7$, and (3) \textit{Kp\textsubscript{max}} $\geq 7$. Within these windows, we align them such that the maximum \textit{Kp} is at the two-day mark. 

\begin{figure}[htb!]
	\centering
	\small
	\includegraphics[width=\textwidth]{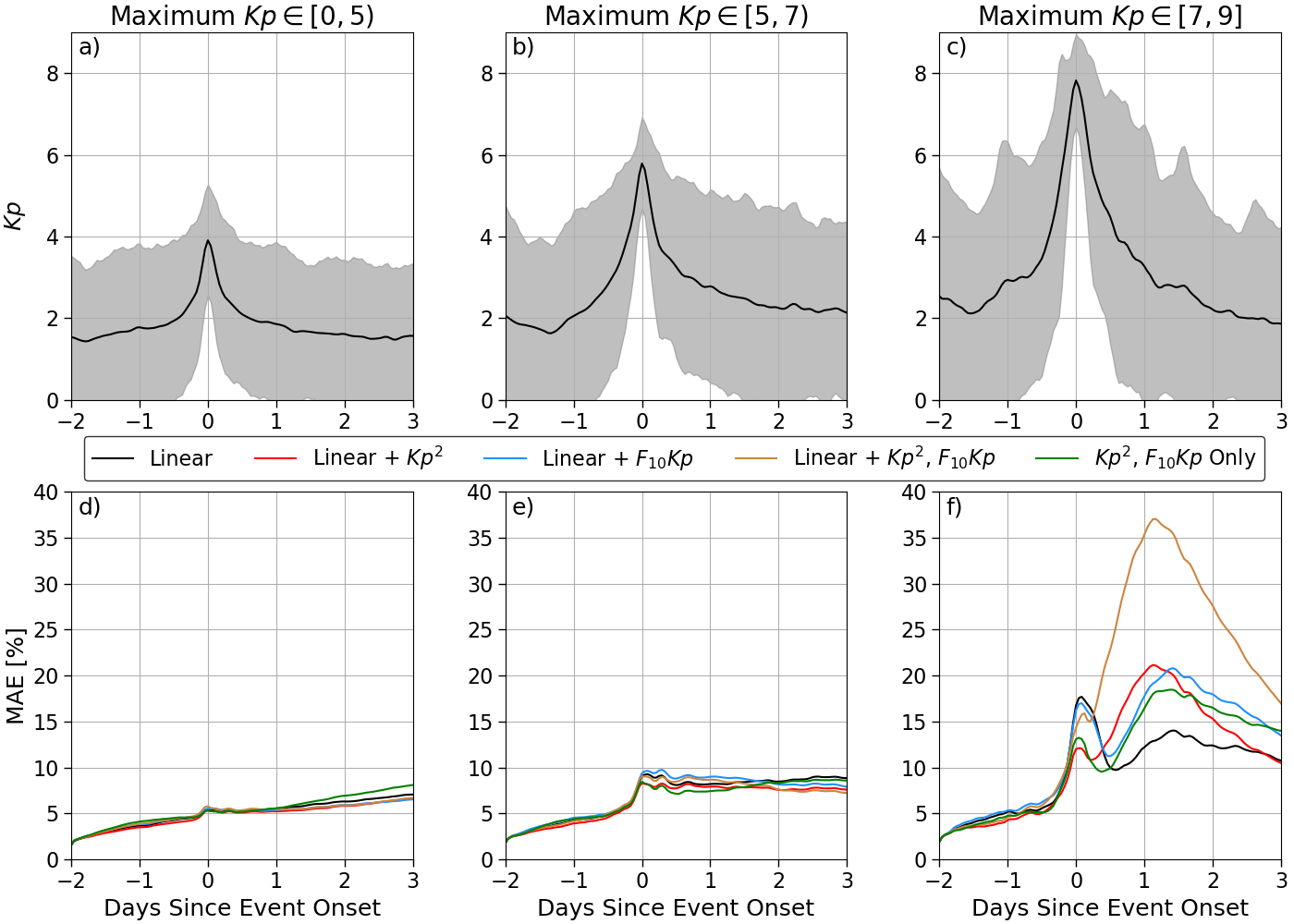}
	\caption{Average \textit{Kp} for the three conditions with shaded $2\sigma$ bounds (a--c) with the corresponding errors (d--f). The legend denotes the models used in panels (d--f). Note: all models use the same temporal inputs.}
	\label{f:DMD_mae}
\end{figure}

We then create DMDc models with five different input sets based on TIE-GCM from 2002--2008 (the LSTM training set). All models use $t_1$--$t_4$ as temporal inputs (see Equation \ref{eqT}). The linear inputs are \textit{F\textsubscript{10}} and \textit{Kp}, and the two nonlinear inputs are the ones used by Gondelach and Linares. Figure \ref{f:DMD_mae} shows the mean absolute error as a function of time for these models.

For the quiet case, panels (a,d), all DMDc models have similar errors. The average error across the five-day windows are within 0.6\%. Although the periods are aligned with respect to the maximum \textit{Kp}, the level of geomagnetic activity is low, and the error is therefore minimally affected. For the moderate case, panels (b,e), there is a sharper rise in error around the onset of a storm -- nearly doubling in $\sim$12 hours. Still, there is little deviation between the models, and the mean error for all models is within 0.7\% over five days. 

The strong storms create stark differences in model performance. While the strictly linear approach has the highest errors at maximum \textit{Kp}, it has the best recovery from the storm. Including the additional nonlinear inputs results in the worst post-storm performance which could be a result of using too many drivers, all changing drastically during these periods. Using only the two nonlinear drivers (and time) does reduce the error at maximum \textit{Kp} by about 5\%, and the rise in error after the storm is not as pronounced. We proceed with the use of this DMDc approach due to storm-time improvement and use in other work. For Figures \ref{f:Sim1_5day_pca}, \ref{f:Sim1_year_density}, and \ref{f:Orbit} in the proceeding section, we show both the linear and nonlinear-input DMDc models which will be referred to as DMDc and DMDc NL, respectively. All other tables and figures use the nonlinear-input DMDc model.

\section{Results}\label{sec:results}

This dynamic reduced order modeling effort poses challenges when determining the best way to evaluate the models. With static modeling (e.g. HASDM-ML \cite{HASDM_ML}, CHAMP-ML\cite{UQtech}, MSIS-UQ \cite{msis_uq}), error and calibration can be analyzed relatively simply across the training, validation, and test sets. However, the DMDc and LSTM ensemble models will have different statistics depending on the evaluation window. We therefore must take careful consideration when evaluating and comparing the two methods.

\subsection{Five-Day Operational Analysis}\label{sec:LSTM5day}

Once the final models were trained and the weighting and scaling schemes were determined (Section \ref{sec:ensemble_scaling}), the ensemble was evaluated on all available TIE-GCM data. The results for five-day dynamic prediction windows on the training, validation, and test sets is shown in Table \ref{t:LSTM_5-day} alongside the DMDc model. The calibration error score (Equation \ref{eqces}) is shown for TIE-GCM ROPE.

\begin{table}[htb!]
	\fontsize{10}{10}\selectfont
    \caption{Error and calibration statistics for DMDc and LSTM models averaged over 5-day dynamic prediction periods.}
   \label{t:LSTM_5-day}
        \centering
   \begin{tabular}{| c | c | c | c | c | c | c | c | c | c |} 
      \hline 
        \multirow{6}{*}{\textbf{DMDc}} & \textbf{Set} & \multicolumn{7}{c|}{\textbf{Training}} \\ \cline{2-9}
        & \textbf{Year} & \textbf{2002} & \textbf{2003} & \textbf{2004} & \textbf{2005} & \textbf{2006} & \textbf{2007} & \textbf{2008} \\ \cline{2-9}
        & \textbf{MAE} & 6.31\% & 7.55\% & 6.81\% & 6.41\% & 6.02\% & 5.70\% & 5.97\% \\ \cline{2-9}
         & \textbf{Set} & \textbf{Val.} & \multicolumn{6}{c|}{\textbf{Test}} \\ \cline{2-9}
         & \textbf{Year} & \textbf{Sim1} & \textbf{1996} & \textbf{1997} & \textbf{1998} & \textbf{1999} & \textbf{2000} & \textbf{2001} \\ \cline{2-9}
        & \textbf{MAE} & 32.43\% & 5.00\% & 4.48\% & 5.43\% & 6.73\% & 7.01\% & 6.34\% \\ \hline\hline
        
        \multirow{8}{*}{\textbf{LSTM}} & \textbf{Set} & \multicolumn{7}{c|}{\textbf{Training}} \\ \cline{2-9}
        & \textbf{Year} & \textbf{2002} & \textbf{2003} & \textbf{2004} & \textbf{2005} & \textbf{2006} & \textbf{2007} & \textbf{2008} \\ \cline{2-9}
        & \textbf{MAE} & 5.66\% & 6.44\% & 5.93\% & 6.87\% & 5.12\% & 7.23\% & 8.45\% \\ \cline{2-9}
        & \textbf{CES} & 16.16\% & 16.83\% & 15.08\% & 15.75\% & 15.56\% & 16.02\% & 18.49\% \\ \cline{2-9}
        & \textbf{Set} & \textbf{Val.} & \multicolumn{6}{c|}{\textbf{Test}} \\ \cline{2-9}
        & \textbf{Year} & \textbf{Sim1} & \textbf{1996} & \textbf{1997} & \textbf{1998} & \textbf{1999} & \textbf{2000} & \textbf{2001} \\ \cline{2-9}
        & \textbf{MAE} & 10.34\% & 5.57\% & 4.83\% & 5.60\% & 6.27\% & 6.38\% & 6.44\% \\ \cline{2-9}
        & \textbf{CES} & 13.53\% & 17.60\% & 15.13\% & 15.03\% & 15.16\% & 14.82\% & 15.90\% \\ 
      \hline
   \end{tabular}
\end{table}

Across the training set (decline of solar cycle 23), both DMDc and the ensemble have similar errors, on the order of 4--9\%. It is important to note that the errors are with respect to the TIE-GCM density, so it also includes an average of 2\% truncation error from PCA. On the validation set, both modeling approaches have their highest error -- 32\% and 10\% for DMDc and LSTM ensemble, respectively. This is caused by the variability in the Sim1 dataset (Figure \ref{f:LSTM_Sim1_Tuner}). There is a high concentration of storms, and solar activity changes the drastically on the time-scale of days. This causes some issues for the LSTM, an increase of about 4\% error with respect to its average on the training set, but it causes the average DMDc error to jump 4--5 times its normal values for a given year. This will be explored further in Section \ref{sec:LSTM_kp}. 

\begin{figure}[htb!]
	\centering
	\small
	\includegraphics[width=\textwidth]{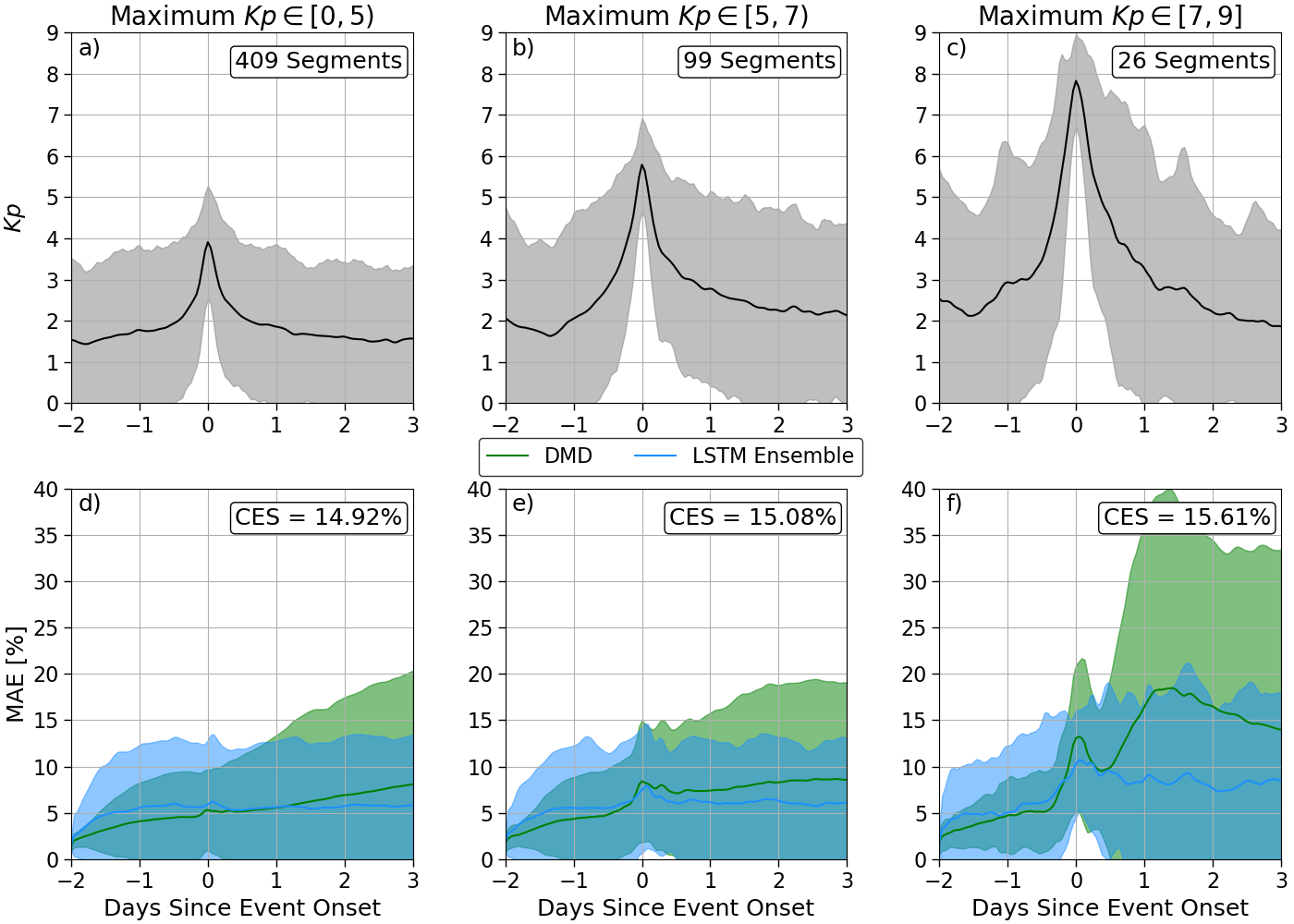}
	\caption{Average \textit{Kp} for the three conditions (a--c) with the corresponding errors for DMDc and LSTM ensemble(d--f). The shading represents $2\sigma$ bounds for \textit{Kp} (a--c) and errors (d--f).}
	\label{f:Kp_ensemble_mae}
\end{figure}

On the test set, both models perform similarly to both each other and to their performance on the training set. This indicates good generalization on historical periods. The calibration error score for the ensemble is between 13\% and 19\% for any given year (including the validation set). While this is higher than in previous modeling efforts, dynamic prediction poses a challenge. Adding models and architectures to the ensemble could potentially reduce the CES. We test the robustness of the LSTM ensemble to geomagnetic activity by performing the study for DMDc (Figure \ref{f:DMD_mae}) on TIE-GCM ROPE. The results are shown in Figure \ref{f:Kp_ensemble_mae}.

As seen in Figure \ref{f:DMD_mae}, DMDc has low dynamic prediction error during geomagnetically quiet conditions. For this same period, the LSTM ensemble also has low errors, but the error climbs to $\sim5\%$ within 24 hours while it takes DMDc around 72 hours to reach the same error. Beyond this point, the LSTM ensemble has lower errors. In panel (e), the error again climbs to 5\% for the LSTM in the first day, but it is relatively unaffected by the onset of the storms. DMDc errors jump to around 8\% and slightly increases post-storm while the ensemble only jumps to approximately 7\% and drops to around 6\% for the remainder of the dynamic prediction window. 

For the strongest storms (panel (f)), TIE-GCM ROPE proves to be much more robust. The errors for it and DMDc converge around 24 hours after the dynamic prediction starts. At maximum \textit{Kp}, the error for TIE-GCM ROPE peaks at just above 10\% which is below even the lowest error at max \textit{Kp} for any nonlinear DMDc approach tested in Figure \ref{f:DMD_mae}. The LSTM error also continues to decrease post-storm back to the 7\%--9\% range. The $2\sigma$ error bounds to TIE-GCM ROPE area also considerably lower than for DMDc for the storm and post-storm periods. The calibration error score for TIE-GCM ROPE is also fairly consistent across the three conditions.

\subsubsection{DMDc Sensitivity}\label{sec:LSTM_kp}

\begin{figure}[htb!]
	\centering
	\small
	\includegraphics[width=\textwidth]{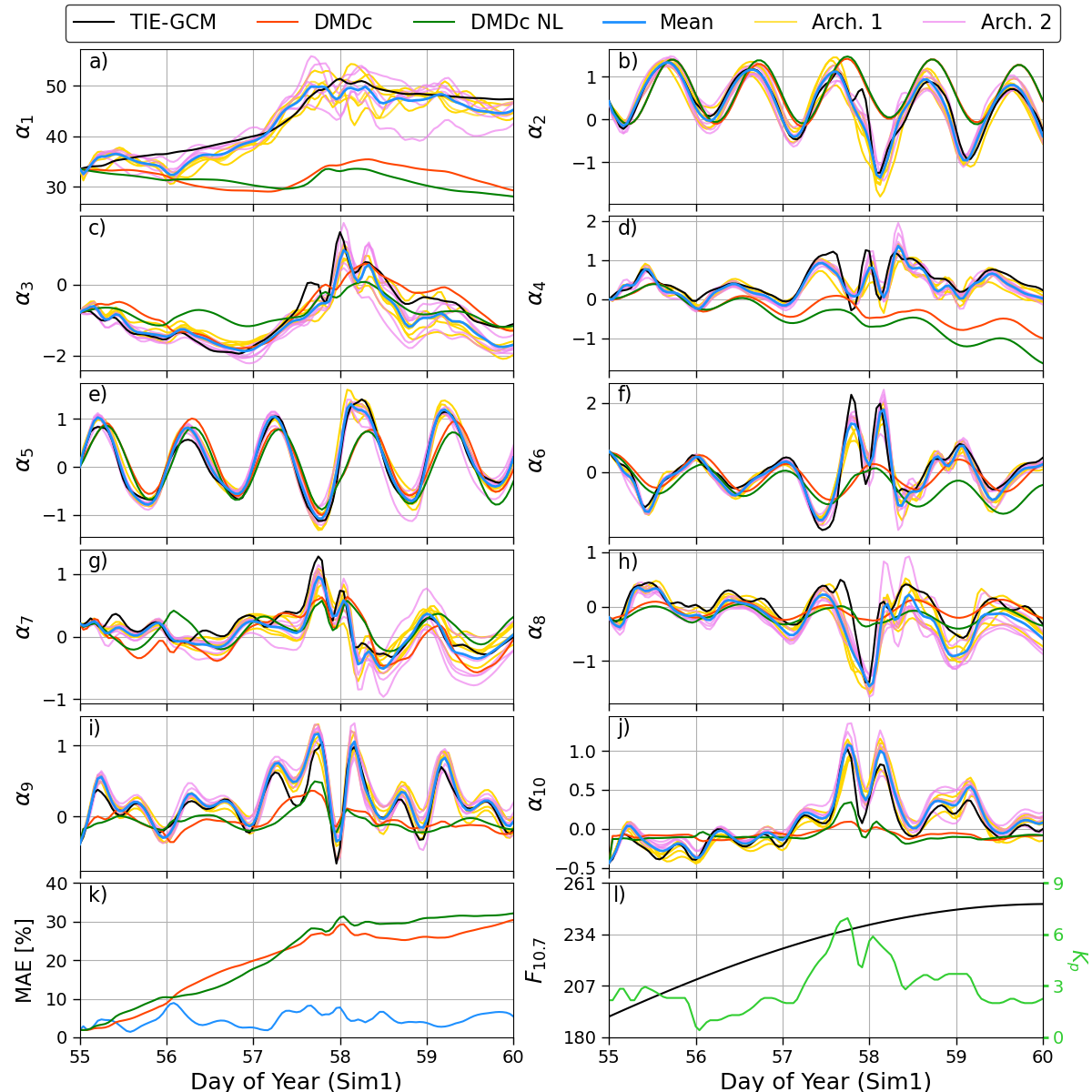}
	\caption{PCA coefficients from TIE-GCM along with dynamic prediction from the linear-input DMDc model, the nonlinear-input DMDc model (DMDc NL) and TIE-GCM ROPE (a--j), global density mean absolute errors for the two models (k), and corresponding space weather drivers (l).}
	\label{f:Sim1_5day_pca}
\end{figure}

Figure \ref{f:Kp_ensemble_mae} highlighted the robustness of TIE-GCM ROPE to geomagnetic activity while also showing the low error for DMDc during quiet periods. However, we noted in Table \ref{t:LSTM_5-day} that the five-day forecast errors for DMDc were significantly higher on the Sim1 validation data. While Sim1 contains an above-average number of geomagnetic storms for a year-long period, another distinguishing feature of the dataset is how drastically \textit{F\textsubscript{10}} can vary in a few days. We explore this further by considering a period from Sim1 that is outside the validation set. The DMDc and TIE-GCM ROPE predictions in the reduced state are displayed in Figure \ref{f:Sim1_5day_pca}. Note: although we extract $\sigma$ from the LSTMs to get distribution statistics for TIE-GCM ROPE, the individual model predictions are shown alongside the ensemble mean for observational purposes. We also show the linear-input DMDc model for comparison.

Before discussing the DMDc and LSTM predictions in Figure \ref{f:Sim1_5day_pca}, we must look at \textit{F\textsubscript{10}} and \textit{Kp} for this period (panel (l)). \textit{F\textsubscript{10}} increases from 190--250 sfu in this five-day window. This is a substantial increase for this short of a time -- which is very unlikely -- but it provides us with a test for how well these models can function as a reduced order model for TIE-GCM. \textit{Kp} is fairly quiet for the beginning of this period but rises to a moderate storm just before the three-day mark. 

The most glaring result from Figure \ref{f:Sim1_5day_pca} is the DMDc predictions in panel (a). While the individual LSTMs properly track the quickly rising $\alpha_1$ from TIE-GCM, the two DMDc models do not. In fact, it decreases until the storm onset at day 57. Considering that $\alpha_1$ is the most important PCA coefficient (capturing the most variance in the dataset), this explains the larger DMDc errors in panel (k). While DMDc has generally low errors (Table \ref{t:LSTM_5-day} and Figure \ref{f:Kp_ensemble_mae} panel (d)) it does not seem to follow large variations in the thermosphere, which is a major attribute of the Sim1 dataset. Conversely, TIE-GCM ROPE does well following the variations in all of the PCA coefficients, and the onset of the storm leads to some divergence in the individual LSTM predictions. This is expected as the variance should rise with storms, especially considering that it does affect the model performance (Figure \ref{f:Kp_ensemble_mae}). The ability for TIE-GCM ROPE to follow even the higher-order PCA coefficients shows that it can represent more of the dynamics in TIE-GCM relative to DMDc. It is also important to note how the distributions for each architecture evolve in different ways. For $\alpha_1$ specifically, the second architecture results in more diverse model predictions following the storm, highlighting the importance of the hierarchical development of the ensemble.

In SM1, we show a video of global density maps at 400 km across the 6-day period encompassing the 2003 Halloween storm. These maps were generated for TIE-GCM, TIE-GCM ROPE, the linear-input DMDc model, and the nonlinear-input DMDc model. The diurnal structure is present in all four models for the entire pre-storm period. At the onset of the first storm, the high-density regions for TIE-GCM appear at the poles, move towards the equator, then spread longitudinally. This happens twice in the first day of the storm. The general structure of the thermosphere is highly perturbed by both storms in this period, and there is considerable lateral movement of the high-density regions. Most of this behavior is also present in TIE-GCM ROPE but not in the DMDc models. While the nonlinear-input DMDc model does not show the same movement as TIE-GCM, it shows a considerable improvement over the linear-input model. One area for improvement in TIE-GCM ROPE during this period is the modeling of longitudinal movement of the high-density regions. This limitation may stem from the use of PCA.

\vspace{4mm}

\subsection{Ensemble Emulation}\label{sec:LSTMemulation}

To test emulation capabilities, the DMDc model and TIE-GCM ROPE were evaluated on the same periods from Section \ref{sec:LSTM5day}, but with a dynamic prediction window of approximately one year. The results are displayed in Table \ref{t:LSTM_year}. Note that for 1996, there is only 280 days available for prediction, and the validation period on Sim1 is only 54 days.

\vspace{3mm}

\begin{table}[htb!]
	\fontsize{10}{10}\selectfont
    \caption{Error and calibration statistics for DMDc and LSTM models averaged over full-length dynamic prediction periods. This is 280 days for 1996, 362 days for all other years, and 52 days for the validation set.}
   \label{t:LSTM_year}
        \centering
   \begin{tabular}{| c | c | c | c | c | c | c | c | c | c |} 
      \hline 
        \multirow{6}{*}{\textbf{DMDc}} & \textbf{Set} & \multicolumn{7}{c|}{\textbf{Training}} \\ \cline{2-9}
        & \textbf{Year} & \textbf{2002} & \textbf{2003} & \textbf{2004} & \textbf{2005} & \textbf{2006} & \textbf{2007} & \textbf{2008} \\ \cline{2-9}
        & \textbf{MAE} & 61.94\% & 45.79\% & 49.11\% & 55.90\% & 59.36\% & 75.76\% & 81.30\% \\ \cline{2-9}
         & \textbf{Set} & \textbf{Val.} & \multicolumn{6}{c|}{\textbf{Test}} \\ \cline{2-9}
         & \textbf{Year} & \textbf{Sim1} & \textbf{1996} & \textbf{1997} & \textbf{1998} & \textbf{1999} & \textbf{2000} & \textbf{2001} \\ \cline{2-9}
        & \textbf{MAE} & 44.46\% & 82.02\% & 44.10\% & 40.21\% & 42.90\% & 52.24\% & 38.46\% \\ \hline\hline
        
        \multirow{8}{*}{\textbf{LSTM}} & \textbf{Set} & \multicolumn{7}{c|}{\textbf{Training}} \\ \cline{2-9}
        & \textbf{Year} & \textbf{2002} & \textbf{2003} & \textbf{2004} & \textbf{2005} & \textbf{2006} & \textbf{2007} & \textbf{2008} \\ \cline{2-9}
        & \textbf{MAE} & 6.02\% & 6.75\% & 6.72\% & 8.14\% & 7.81\% & 12.94\% & 23.12\% \\ \cline{2-9}
        & \textbf{CES} & 15.78\% & 24.09\% & 26.48\% & 27.63\% & 28.52\% & 26.82\% & 22.28\% \\ \cline{2-9}
        & \textbf{Set} & \textbf{Val.} & \multicolumn{6}{c|}{\textbf{Test}} \\ \cline{2-9}
        & \textbf{Year} & \textbf{Sim1} & \textbf{1996} & \textbf{1997} & \textbf{1998} & \textbf{1999} & \textbf{2000} & \textbf{2001} \\ \cline{2-9}
        & \textbf{MAE} & 11.27\% & 18.70\% & 9.67\% & 6.21\% & 6.76\% & 6.73\% & 6.76\% \\ \cline{2-9}
        & \textbf{CES} & 5.59\% & 30.06\% & 29.29\% & 25.39\% & 17.30\% & 12.21\% & 15.11\% \\ 
      \hline
   \end{tabular}
\end{table}
\vspace{4mm}

\begin{figure}[htb!]
	\centering
	\small
	\includegraphics[width=\textwidth]{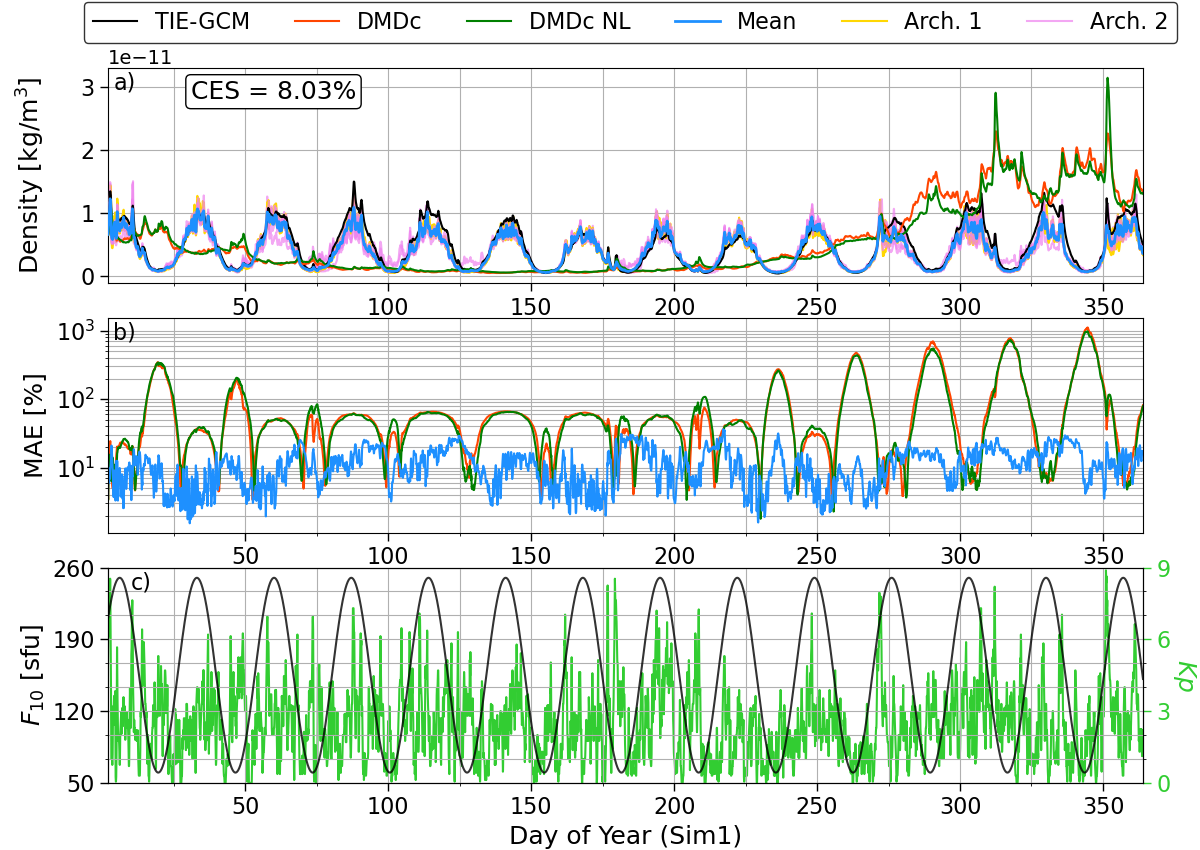}
	\caption{Mean density at 400 km (a) with global-averaged errors for linear-input DMDc, nonlinear-input DMDc (DMDc NL), and LSTM ensemble (b), and the corresponding space weather drivers (c) for a 362-day period across the Sim1 dataset.}
	\label{f:Sim1_year_density}
\end{figure}

For the long-term dynamic prediction, DMDc becomes unreliable with errors ranging from 38\%--82\%. For most years, the TIE-GCM ROPE error is similar to the corresponding values in Table \ref{t:LSTM_5-day}. It appears that the ensemble can emulate TIE-GCM for long periods with low errors with the exception of solar minimum (1996, 2007, 2008). During these periods, the errors are between 13\% and 23\%. The calibration error score for the ensemble is generally higher for these long-term dynamic prediction windows. To visualize the long-term dynamic prediction performance, we look at a 362-day prediction on the Sim1 dataset. The validation set is contained within this period but only accounts for $\sim$$15\%$ of the samples. The mean density at 400 km for TIE-GCM  is shown with DMDc and TIE-GCM ROPE predictions in Figure \ref{f:Sim1_year_density}.

While it is difficult to see, both DMDc and TIE-GCM ROPE have low errors for the first few days, but as seen with Figure \ref{f:Sim1_5day_pca}, the DMDc error quickly compounds. The DMDc predictions do not follow the trends of TIE-GCM, but the cyclic nature of \textit{F\textsubscript{10}} in Sim1 allows for the DMDc error to briefly drop at times (panel (b)). However, the large errors, peaking above 1000\%, require panel (b) to be shown in a logarithmic scale. TIE-GCM ROPE is able to track the long and short-term variations without any state updates. Its errors peaks at 30\% but generally remains around 10\% across the year-long prediction window. The individual LSTM predictions are more prominent during high-density levels in solar maximum conditions (see panel (c)), and the uncertainty is therefore higher in those conditions. 

\vspace{4mm}

\subsection{Implications}\label{sec:orbit}

To further investigate the benefits of the LSTM ensemble approach, we consider a 72-hour satellite state propagation during the major geomagnetic storm on November 20, 2003 for a highly inclined orbit. The initial state and parameters for the ballistic coefficient are shown in Table \ref{t:orbit_params}. The mass and cross-sectional area come from the CHAMP satellite with zero-attitude \cite{CHAMP}. The drag coefficient is assumed to be constant and comes from neural network models from \cite{smriti}. The only perturbations considered in this analysis are two-body, \textit{J\textsubscript{2}}, and atmospheric drag.

\begin{table}[htb!]
	\fontsize{10}{10}\selectfont
    \caption{Initial state in the Cartesian reference frame and satellite parameters.}
  \label{t:orbit_params}
        \centering 
  \begin{tabular}{| c | c | c |} 
    \hline
        \textbf{X [m]} & \textbf{Y [m]} & \textbf{Z [m]} \\ \hline
        \;\;\;\;\;3782900.7032\;\;\;\;\; & \;\;\;\;\;-5441600.6779\;\;\;\;\; & \;\;\;\;\;-1420075.1327\;\;\;\;\; \\ \hline
        \textbf{V\textsubscript{X} [m/s]} & \textbf{V\textsubscript{Y} [m/s]} & \textbf{V\textsubscript{Z} [m/s]} \\ \hline
        -606.6600 & 1539.2559 & -7488.3946 \\ \hline
        \textbf{C\textsubscript{D}} & \textbf{A [m\textsuperscript{2}]} & \textbf{m [kg]} \\ \hline
        3.0912 & 0.7710 & 500 \\
    \hline
  \end{tabular}
\end{table}

\begin{figure}[htb!]
	\centering
	\small
	\includegraphics[width=\textwidth]{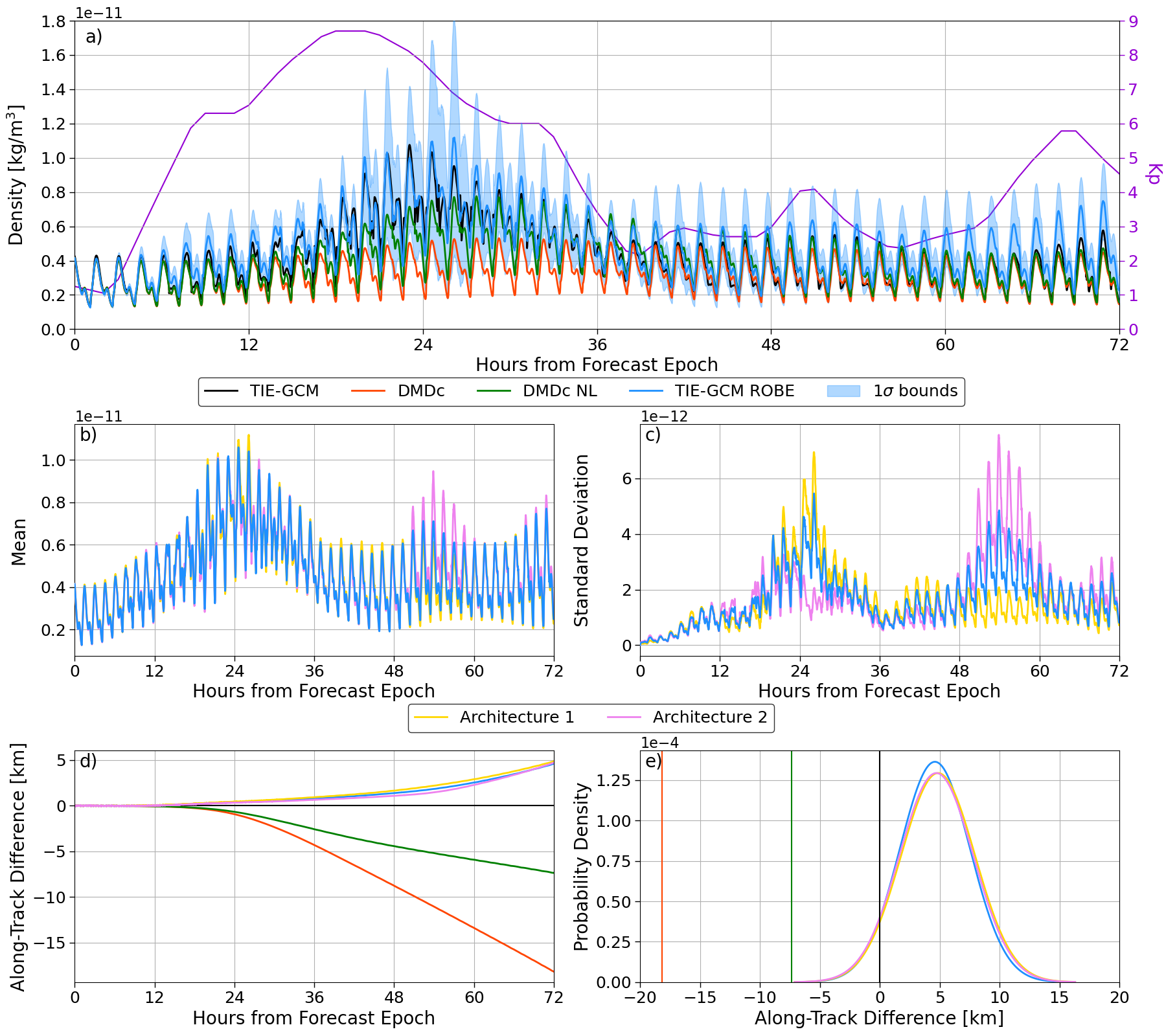}
	\caption{Density along the orbit for TIE-GCM and all modeling approaches with \textit{Kp} (a), mean density (b) and standard deviation (c) predictions for the different architectures, along-track position difference as a function of time (d), and final along-track position difference (e) for the November 2003 geomagnetic storm.}
	\label{f:Orbit}
\end{figure}

For TIE-GCM and the two DMDc approaches discussed in previous sections, we simply propagate the satellite state using the Domand-Prince Runge-Kutta method and interpolate density in space and time from the model \cite{DP}. This results in a single possibility of the satellite state at each point in time. For TIE-GCM ROPE, there are additional considerations as it is a probabilistic model. To account for this, we use a Monte Carlo (MC) method (1,000 samples) with a first-order Gauss-Markov process to sample density using an 18-minute half-life \cite{superensemble}. Since there was variability between the architectures in Figure \ref{f:Sim1_5day_pca}, this MC analysis is also performed using each architecture independently. The results for this study are shown in Figure \ref{f:Orbit}.

Figure \ref{f:Orbit} panel (a) shows the density from TIE-GCM, linear-input DMDc, nonlinear-input DMDc, and TIE-GCM ROPE. For TIE-GCM ROPE, there is a shaded region showing the $1\sigma$ bounds for density from the 1,000 runs. The secondary y-axis shows the corresponding $Kp$ for this period. The TIE-GCM ROPE mean closely follows the rise and fall of density observed by TIE-GCM. Meanwhile, the linear-input DMDc model shows an slight, unobservable increase in density during the storm. The use of nonlinear inputs does improve storm-time modeling in this case. However, it is not able to reach TIE-GCM levels. The uncertainty bounds for TIE-GCM ROPE increase considerably during the storm and decrease for the post-storm period. Note: \textit{Kp} was interpolated in this work to match the cadence of the dataset. 

Panel (b) shows that there is not much difference between the two architectures for the mean density with the exception of a post-storm spike for architecture 2 between 48 and 60 hours. In panel (c), the is a noticeable difference in the density standard deviation between architectures. During the storm, the first architecture has much more variation while the opposite is true after the storm. The hierarchical approach helps combat spikes in uncertainty only originating from a single architecture. 

Panel (d) considers how far the satellite is from the reference point (using TIE-GCM) over the duration of the study. The underestimation of density from the DMDc models is highlighted in how far behind the satellites are relative to TIE-GCM. The models noticeably diverge from TIE-GCM around 18 hours, and their biases accumulate quickly. The nonlinear inputs help decrease the bias by a factor of roughly $1/2$. The LSTMs overpredict density for this period, but their biases are much less pronounced. There is an inflection in the LSTM biases around 60 hours when the models consistently overpredict density relative to TIE-GCM. 

Panel (e) highlights the importance of probabilistic modeling. The two DMDc models are deterministic and their final along-track position differences are therefore point estimates. The use of nonlinear inputs reduces the bias from 18 km to 7.5 km relative to TIE-GCM. TIE-GCM ROPE, being probabilistic, provides a distributions of positions as a result of the MC approach. The final position spread from TIE-GCM ROPE is approximately 20 km due to the strength of the storm. However, the bias relative to TIE-GCM is less than 5 km, and the TIE-GCM position is captured by the distribution.

\section{Summary}\label{sec:con}

TIE-GCM ROPE is an ensemble of ten LSTMs trained on seven years of TIE-GCM outputs during solar cycle 23. The direct probability method with NLPD for UQ used in previous work proved to be ineffective for this dynamic modeling application, which spurred an ensemble approach. A fixed weighting scheme was determined based on individual model performance throughout the training set to get a weighted mean prediction based on observed errors. We use the ensemble predictions to extract a sample standard deviation and use a scaling method to ensure better calibration of the uncertainty estimates. This is done separately for the two subsets of LSTM models resulting from different architectures which requires a hierarchical approach.

TIE-GCM ROPE was compared to Dynamic Mode Decomposition with control, a popular dynamic ROM technique for the thermosphere. We show that while general errors are virtually equivalent for the two approaches, TIE-GCM ROPE does considerably better in capturing the dynamics of TIE-GCM. It also follows the data for extreme periods and has the capability to emulate the system, making dynamic predictions (no state update) for up to a year or over 8,700 time steps with approximately 10\% mean error. We show in SM1 that TIE-GCM ROPE also better models the dynamics in the thermosphere during the 2003 Halloween storm. This is behavior that the DMDc models are not able to capture.

An operational study was conducted to see the effect of different TIE-GCM surrogate models on satellite state propagation during a geomagnetic storm. This showed that nonlinear inputs improved modeling for DMDc, but it was not able to reach the magnitude of density represented by TIE-GCM. This caused a significant position bias to form relative to TIE-GCM. For the nonlinear-input DMDc model, the final along-track position bias was approximately 7.5 km. For TIE-GCM ROPE, we show that the hierarchical approach is necessary as the individual architectures predict diverse possibilities of the density distribution, and the combination of the two negates the effect of one architecture predicting significant uncertainty. TIE-GCM ROPE had a final position bias under 5 km relative to TIE-GCM, but a Monte Carlo analysis was able to capture the position of TIE-GCM while DMDc only provided point estimates.

\section*{Future Work}

This work is meant to lay the foundation for a different approach to reduced order modeling for physics-based thermosphere models. We show that the use of LSTMs provide improvements in both long-term modeling and modeling nonlinear periods (e.g. geomagnetic storms). However, a limitation for this approach is the use of the linear PCA method for dimensionality reduction. Nonlinear techniques can be implemented to further improve storm-time reduced order modeling. The probabilistic approach through hierarchical ensemble modeling is important (Section \ref{sec:orbit}) but could be further improved with the addition of more models and architectures. We will also advance this framework to become a reduced order Bayesian emulator (ROBE) through the introduction of data assimilation.

\section*{Data Availability}

All TIE-GCM and input data used in this work comes from Mehta et al. \cite{MehtaROM}. Details on TIE-GCM are described by \cite{TIEGCM,TIEGCM_orig}. The video for SM1 can be found at \url{https://github.com/rjlicata/tie-gcm_rope_preprint}.

\section*{Acknowledgments}

PMM gratefully acknowledges support under NSF CAREER award \#2140204. The authors would like to thank Dr. Snehalata Huzurbazar for providing insight to hierarchical ensemble modeling.

\bibliographystyle{ieeetr}  
\bibliography{references}

\end{document}